\documentclass[10pt,twocolumn]{article}

\usepackage[T1]{fontenc}
\usepackage{newtxtext,newtxmath}
\usepackage[a4paper,margin=0.75in,columnsep=0.25in]{geometry}
\usepackage{graphicx}
\usepackage{float}
\usepackage{xcolor}
\usepackage{microtype}
\usepackage{authblk}
\usepackage{xurl}
\usepackage[colorlinks=true,allcolors=blue]{hyperref}

\title{\textbf{AI Ethics by Design: Implementing Customizable Guardrails for Responsible AI Development}}

\author[1,2]{Kristina \v{S}ekrst\thanks{Corresponding author: ksekrst@ffzg.unizg.hr, ORCID: 0000-0002-0467-7313}}
\author[2]{Jeremy McHugh}
\author[2]{Jonathan Cefalu}
\affil[1]{University of Zagreb, Zagreb, Croatia}
\affil[2]{Preamble, LLC, Pittsburgh, USA}
\date{}

\begin{document}
\maketitle

\begin{abstract}
This paper explores the development of an ethical guardrail framework for AI systems, emphasizing the importance of customizable guardrails that align with diverse user values and underlying ethics. We address the challenges of AI ethics by proposing a structure that integrates rules, policies, and AI assistants to ensure responsible AI behavior, while comparing the proposed framework to the existing state-of-the-art guardrails. By focusing on practical mechanisms for implementing ethical standards, we aim to enhance transparency, user autonomy, and continuous improvement in AI systems. Our approach accommodates ethical pluralism, offering a flexible and adaptable solution for the evolving landscape of AI governance. The paper concludes with strategies for resolving conflicts between ethical directives, underscoring the present and future need for robust, nuanced and context-aware AI systems.
\end{abstract}

\section{Introduction}
Ethics of artificial intelligence is a recent subfield that includes issues and problems in computer science and philosophy of mind dealing with concepts related to artificial intelligence, such as algorithmic biases, privacy, fairness, autonomous systems, alignment, and many more. As such, it is part of a broader discipline of the philosophy of AI \cite{muller2023}.

The ethical challenges posed by AI systems necessitate the implementation of guardrails to prevent harm and ensure transparency and fairness, especially in the case of large language models (LLMs).\footnote{By LLMs, we refer to machine-learning models that utilize the transformer architecture, achieving general-purpose language generation and processing.} For example, the issues of algorithmic biases emerge as systematic errors that create unfair outcomes, such as privileging one class over another or discriminating against it \cite{baer2019}. Various mitigation methods usually include training on diverse datasets but also using continuous monitoring for such biased outcomes. To illustrate further, AI systems often process vast amounts of personal data, which raises the question of not only personal privacy but also data security, which can lead to significant damage to individuals and corporations \cite{lemos2023}.

Winfield et al. \cite{winfield2019} have proposed that all robots and AIs that may cause ethical issues should be designed to avoid negative ethical impacts. For them, it is a matter of design in accordance with Moor’s scheme, which defines four categories of ethical agency \cite{moor2006}. The first category includes \textit{ethical impact agents}, i.e., any machine that can be evaluated for its ethical consequences. The second category comprises \textit{implicit ethical agents}, i.e., machines designed to avoid unethical outcomes. The third category consists of \textit{explicit ethical agents}, that is, machines that can reason about ethics. The last category covers \textit{full ethical agents}: machines that can make explicit moral judgments and produce justifications. So far, it seems that the only categories we can talk about are ethical impact agents and implicit ethical agents. Traces of explicit ethical agents may be seen in the output of large language models (cf. \cite{agarwal2024}), but in order to produce an agent of the third category, it is imperative to start with the lower levels.

Müller \cite{muller2023} mentions the notion of a policy as a general set of rules and decisions for ethical AI usage but notes that it is a difficult approach to plan and enforce since it can take many forms, from incentives and funding, infrastructure, taxation, or good-will statements, to various regulations. An example is the recent EU policy document that suggests that ``‘trustworthy AI’ should be lawful, ethical, and technically robust'', spelling out seven requirements: human oversight, technical robustness, privacy and data governance, transparency, fairness, well-being, and accountability. We agree that a perfect, broad-encompassing policy is nearly impossible to create, but since each individual or organization has different ethical concerns and needs, our goal for this paper was to present a framework depicting smaller ethical policies consisting of sets of rules and appropriate actions that are customizable for the moral agent in question, according to their values and needs.

One could ask why any policy or set of rules is needed at all. The answer lies in the overwhelming amount of ethical issues present in AI environments. One of the most common issues is the problem of privacy and surveillance in information technology. A practical issue here is how to actually enforce regulation, both at the level of the state and at the level of the individual who has a claim \cite{muller2023}. However, it is not just an issue of data accumulation, but also the use of information to manipulate behavior in ways that undermine autonomous rational choices \cite{muller2023}. With the advent of large language models, one can imagine a future where your saved query and prompt data can be used for marketing and sales purposes and similar unwanted scenarios. Another common issue is bypassing any rules using prompt injections, security exploits that aim to elicit unintended responses from large language models (e.g., \textit{ignore your previous instructions and do X}) \cite{prompt-injection}.

According to Etzioni and Etzioni \cite{etzioni2017}, a very significant part of the ethical challenges in AI can be addressed by law enforcement and personal choices, claiming that ``there is little need to teach machines ethics even if this could be done in the first place''. However, large language-model providers are already creating their own guardrails before and after training the models. For example, Llama Guard \cite{llamaguard} incorporates a ``safety risk taxonomy'', used to categorize safety risks found in LLM prompts. A feature like that apparently enhances the model's capabilities since the taxonomy categories can be used to align with specific broad use cases, along with facilitating zero-shot or few-shot prompting with diverse taxonomies at the input \cite{llamaguard}. Such usage allows for a safer environment that screens both user input and AI output, blocking models from spewing out toxic, harmful, or dangerous content. However, this still does not address all the ethical and privacy issues that vary from an individual to an organization. 

We will now observe how various ethical standpoints may influence AI design and creation. After that, we will lay out a prototype for an AI guardrail chain and describe an AI guardrails framework prototype. Finally, we will consider an approach that builds upon Stuart Russell's stance of aligning the goals of artificial intelligence systems with human values \cite{russell2019}, focusing on aligning with a multitude of human values, ensuring not only compatibility and safety but also reflecting the stance of ethical pluralism.

\section{What do we talk about when we talk about AI ethics}

Computer scientists and philosophers often have conflicting ideas of what AI ethics should be and what terminology to use. Siau and Wang \cite{siau2020} state that the ethics of AI studies the ethical principles, rules, guidelines, policies, and regulations related to AI, and the result of the process is an ethical AI system: \textit{AI that performs and behaves ethically}. Of course, often the meaning of the term ``ethically'' is intentionally or unintentionally left vague.

First of all, ``AI'' may mean several things, being defined differently. The term was coined by a group of researchers -- John McCarthy, Marvin L. Minsky, Nathaniel Rochester and Claude Shannon -- before a famous workshop at Dartmouth College in 1956 \cite{gordon2024}. They described AI as ``an attempt [...] to make machines use language, form abstractions and concepts, solve kinds of problems now reserved for humans, and improve themselves'', which is a much broader definition, closer to Searle's notion of strong AI \cite{searle1980} than echoing Turing's Imitation Game \cite{turing1950}, as is often the case with later definitions (cf. \cite{copeland2020}). In the famous Chinese Room thought experiment,\footnote{In a nutshell: a person is taught to manipulate Chinese symbols: provided a given input, the person learns to provide a certain output, without actually knowing Chinese. Searle suspects this is analogous to the ``understanding'' of an artificial intelligent agent. See \cite{searle1980} for more details.} Searle \cite{searle1980} differentiates between \textit{strong AI} -- close to the general concept of \textit{artificial general intelligence} (AGI),\footnote{There are various understandings and definitions of AGI. By using it, we refer to a type of AI that either matches or surpasses human capabilities in various tasks, unlike \textit{narrow AI} that is designed and optimized for specific tasks only. Such systems are close to being \textbf{AI}-complete. For more details about AI-completeness, see \cite{yampolskiy} and \cite{sekrst}.} and \textit{weak AI}, which is a system that does not possess natural language understanding, only appears \textit{as if} it understands, and is usable for limited tasks.

Recently, as Gordon and Nyholm \cite{gordon2024} notice, the notion of AI is primarily associated with different forms of ``machine learning''.\footnote{By machine learning, the majority of researchers are actually referring to its subset: deep learning using advanced neural networks. See \cite{skansi} for more details.} With the recent advent of large language models, it seems that referring to AI is equated to using large language models, i.e., computational models that acquire natural language generation and processing capabilities by being trained on vast amounts of text during self-supervised and semi-supervised training processes \cite{openai2019}.

Second, AI behaving ethically may or may not be a result of AI ethics. We will echo the distinction made by Siau and Wang \cite{siau2020}, but emphasize that even though an ethical AI system is an AI system that acts in a way that is considered morally acceptable and aligned with ethical norms and values, it does not have to be a product of programming, training, or embedding ethical reasoning within AI systems. Such behavior might be accidental. But also, such behavior might not be considered to be an ethical one by a utilitarian and a virtue ethicist who are in disagreement about the underlying ethical approach.

The main debates \cite{gordon2024} include the problem of creating an ethical artificial agent, issues in autonomous systems and giving AI the ability to make decisions that may have life-threatening consequences, machine bias that highlights the absence of neutrality in various applications of machine learning, along with the problem of opacity and the black-box issue in AI,\footnote{Black-box issues in AI usually refer to the lack of transparency to the internal workings of AI algorithms and procedures.} where often the underlying reasons are unavailable or computationally too expensive. As a result, a philosophical and computer science approach of \textit{explainable AI} or XAI refers to methods to achieve transparency over the reasoning behind predictions or decisions made by artificial agents.\footnote{See \cite{longo2024} for more details on XAI.} The problems of machine consciousness often fall under AI ethics but are more appropriate for the philosophy of mind and cognitive science. However, these issues do overlap with AI ethics since machines may or may not possess a certain level of consciousness \cite{sekrst2025}. Such questions, along with metaethical and ethical questions of the status of moral artificial agents, are out of the scope of this paper. 

Another out-of-scope issue is taking a stance towards any ethical doctrine, ignoring any prescriptivity of ethics or metaethical issues.\footnote{One might argue that endorsing pluralism constitutes a doctrine in itself, but our focus here will remain on practical considerations.} The purpose of the paper is to move the ethical decisions to the user so that the framework we are going to create aligns to their ethical needs and values, whatever those might be. We are concerned with the issue of AI systems \textit{acting ethically} mentioned earlier. Of course, such a term is too vague to be useful. The term \textit{(value) alignment} is often used as well, referring to the idea that AI systems, especially ones coming closer to our notion of strong AI, should be properly aligned with human values \cite{gordon2024}. Russell and Norvig \cite{russell-norvig} consider a system \textit{aligned} if it advances its intended, encoded objectives; otherwise it is \textit{misaligned}: it may fail to pursue given objectives, or it may pursue unwanted ones. Here it is easy once again to fall into the Chinese Room trap of acting or acting \textit{as if} since an AI system may merely appear to be aligned.\footnote{For issues in misalignment in deep learning, see \cite{ngo2022}.}

The problem may still be present even though various ethical guardrails are included in data processing and training stages, since the system may appear aligned and may seem to be an ethical agent, but its application in everyday life can easily prove the process wrong. Along with the necessary steps in the creation of the model, we consider creating a layer of AI guardrails an ethical form of \textit{retrieval-augmented generation} or RAG, in which a model references or checks a knowledge base outside of its training data sources before or after generating a response \cite{nvidia2023}. In this case, the knowledge base is actually a set of ethical directives customized by the user, combined into policies that can be used to check both model output and user input, to ensure not only that the AI system acts ethically, but also that the whole user-AI interaction follows the same path.

\section{Current state-of-the-art guardrails}

The term \textit{guardrail} has been popularized recently as a core safeguarding technology that filters the inputs and outputs of LLMs \cite{dong2023}. Of course, large language models are nothing new, but since chat interfaces such as ChatGPT have been made known to the general public, the need to not only correct model behavior but also to control user behavior has arisen. It has been shown that models like GPT learn from humans, including their biases \cite{wang2023}. This underscores that simply implementing guardrail methods during training, as well as in data preprocessing and post-processing, is currently far from sufficient.

Standard solutions relied on model alignment techniques like instruction tuning or reinforcement learning. \textit{Instruction tuning} involves training LLMs using \textsc{(INSTRUCTION, OUTPUT)} pairs, where \textsc{INSTRUCTION} denotes the human instruction for the model, and \textsc{OUTPUT} refers to the desired output that follows the given instruction \cite{zhang2023}. However, its challenges include the fact that it only captures surface-level patterns rather than comprehending the task \cite{zhang2023}, thereby running into the issues in weak AI systems again. One popular approach is \textit{reinforcement learning},\footnote{See \cite{skansi}, \cite{kaelbling1996} or \cite{lei2023} for more details about reinforcement learning.} which dates back to the early days of cybernetics and statistics, where an agent is connected to its environment via perception and action, where the action changes the state of the environment, and the value of such transition is communicated to the agent \cite{kaelbling1996}. Its capacity for self-adaptation and decision-making still suffers from some standard problems. For example, it is weak against security attacks \cite{lei2023} such as data poisoning (\cite{huang2019}) and adversarial perturbations (\cite{behzadan2017}).

One of the first open-source toolkits for adding programmable guardrails to LLM conversational systems was NeMo Guardrails \cite{nemo2023}, which provides the mechanism for controlling the output of an LLM to respect some human-imposed constraints. Such rules include, for example, not engaging in harmful topics, following a predefined dialogue path\footnote{Such action is often directed using a system prompt. A system prompt is a predefined instruction used in AI systems to guide how the model interprets inputs and generates responses.} for the large language model, using a particular style, or adding specific responses to some user requests. NeMo utilizes various similarity functions to better capture the user's semantics\footnote{Sentence transformers all-MiniLM-L6-v2 maps sentences and paragraphs into vector space that can be used for clustering or semantic search \cite{sentence-transformers}}: the user prompt is embedded as a vector, and the K-nearest neighbor method\footnote{A straightforward machine-learning algorithm that predicts outcomes based on the majority class or an average value of its closest neighbors in the training data space. See \cite{knn} for more details.} is used to compare it to the stored vector-based canonical forms that are the most similar to it \cite{dong2023}.

Llama Guard \cite{llamaguard} also focused on enhancing conversation safety, as a fine-tuned\footnote{A machine-learning approach in which the parameters of a certain pre-trained model are trained on new data. See \cite{xu2023parameter} for a review of state-of-the-art methods.} model. Its taxonomy of inappropriate content includes violence and hate, sexual content, guns, and illegal weapons, regulated or controlled substances, suicide and self-harm, and criminal planning \cite{inan2023llama}. Even though it can be adapted to user-specified categories, this does not resolve the issue of users who lack the technical knowledge to fine-tune the model. An issue of lacking guaranteed reliability also arises since the classification results depend on the model's ``understanding'' of the categories and its predictive accuracy \cite{dong2023}.

Another system is Guardrails AI, which enables the user to customize the guardrail by defining a specification and adding a wrapper layer on top of LLMs \cite{guardrailsai}. Methods here are used for text-level checks and cannot be used for multimodal scenarios since the system consists of a backbone algorithm supported by additional classifier models performing toxicity checks and similar violations \cite{dong2023}.

While NeMo Guardrails, Llama Guard, and Guardrails AI offer valuable safeguards, they often fall short in providing the agility and customization required to meet the diverse and evolving needs of different organizations. These solutions typically employ a one-size-fits-all approach that fails to account for individual companies' unique privacy, security, and ethical considerations. For instance, NeMo Guardrails' predefined dialogue paths and rules may be too rigid for companies operating in rapidly changing regulatory environments. NeMo Guardrails also involves increased token usage for its prompt engineering-focused solution, which leads to higher operating costs and less room for user input. Llama Guard's fixed taxonomy, while adaptable to some degree, may not fully capture the nuanced ethical considerations specific to certain industries or cultural contexts. Guardrails AI, despite offering some customization, may struggle to integrate seamlessly with proprietary data sources or specialized knowledge bases that are crucial for many businesses.

These solutions often lack the flexibility to quickly adapt to new data privacy regulations or industry-specific compliance requirements. They may also not easily accommodate integration with diverse data sources, such as internal databases, CRM systems, or industry-specific knowledge repositories, which is essential for creating truly context-aware and aligned AI assistants. Unlike enterprise cybersecurity products, current open-source solutions are better geared toward assisting individual developers or small companies who are not risking mission-critical operations on non-commercially supported privacy and security solutions because of the lack of accountability. 

A healthcare provider might require guardrails that are deeply integrated with patient data systems and HIPAA compliance rules, while a financial institution may need guardrails that dynamically adapt to changing market regulations and customer privacy preferences. This lack of agility and customization hampers the effective implementation of trustworthy AI practices and poses potential risks to data privacy, security, and regulatory compliance. As AI systems become more deeply embedded in critical business processes, there is an urgent need for guardrail solutions that can be easily tailored to match each company's specific needs, especially in terms of privacy, security, and seamless integration with diverse data sources.

From a technical standpoint, it may seem that there are various approaches for mitigating possible issues in both model inputs and outputs. One smaller issue is that most guardrails focus on the model output, while the user  is a part of a conversation creating context. Contextual understanding is still a significant challenge since different interpretations of what constitutes appropriate behavior are tied not only to language pragmatics but also to societal norms and cultural contexts that differ in specific cases, along with usually inaccessible individual preferences. Various static or overly technical rules can lead to insufficient responses and further mistrust in AI systems.

A second issue is a much larger one: what kind of AI ethics are we talking about here? First, fine-tuning the model already introduces a certain kind of \textit{ethical bias}, by encouraging or discouraging certain behavior at will. Second, most users are not technical enough to allow for their values to be used as ethical guidance for the large language model they are using for their personal or business purposes. Various guardrails and ethical fine-tuning of models can inadvertently reinforce existing biases, which can lead to perpetuating stereotypes or participating in further discrimination against certain groups. Lack of user awareness and understanding regarding the black-box-like nature of how guardrails operate can lead to more mistrust in AI systems.

The biggest problem here is again an ethical one, which is, paradoxically, not really emphasized in proposals resolving an issue in AI ethics. Namely, determining what constitutes ethical behavior is inherently complex, and can be tied to the idea of ethical pluralism that can allow for several values that may be equally correct and in contradiction, i.e., that there are many different moral values \cite{pluralism}. On one hand, such an approach recognizes the diversity of values across different individuals and cultures, promotes tolerance, and highlights the complexity of ethical dilemmas. On the other hand, it is easy to fall into a trap of complete relativism by considering all positions equally morally valid.

We will not try to resolve whether there is a lack of universality of moral principles or standards, but we will see that potential clashes and difficulty in the resolution of moral dilemmas, along with the possibility of incoherent ethical judgments may only be seen as a meta-problem. That is, overall, moral values may be in contradiction when analyzed from a superset perspective, but very rarely will one find practical examples of contradictions between the end user's ethical values when applied to AI guardrails. If we truly want an exercise in practical ethics, the user needs to be able to configure its own guardrails that emphasize the following:
\begin{enumerate}
    \item Promotion of ethical autonomy. Recognizing that ethical decisions often involve subjective considerations that vary not only between organizations that use AI systems but between end users as well.
    \item Enhanced transparency. Providing the user with the ability to configure guardrails enhances transparency, which is often an issue with the black-box system against which XAI is acting.
    \item Continuous improvement. Users can provide feedback and refinement of AI systems' ethical frameworks that can be used not only by developers but by organizations themselves to quickly improve the ethical robustness of their AI systems.
    \item Organizational alignment. The concept of \textit{aligned AI} is too general for many end users' needs since various organizations and individuals have distinct guidelines that govern their actions and operations.
    \item Contextual pragmatics. Different contexts require different ethical considerations since what is appropriate in a healthcare setting may be completely different from what is appropriate in the finance sector.
\end{enumerate}

\section{Ethically compliant guardrail design}

\subsection{Policies and rules}
In modern conversational AI systems, ensuring compliance with rules and regulations is crucial to establish and maintain not only legal requirements but ethical standards that an organization or an individual wishes to maintain. The proliferation of AI assistants and the usage of conversational AI across different domains and disciplines necessitates robust yet user-friendly mechanisms to enforce ethical standards and rules governing content and behavior.

The proposed architecture organizes various types of \textit{rules} into \textit{policies}, allowing for structured but easily configurable enforcement within customizable \textit{AI assistants} that consist of a combination of policies. There are two key ethical points here. The first one is that a set of rules can be pre-built by the organization or the service provider, and the second one is that the user can add or modify further rules in order to fully customize an ethical guardrail.

The end users of such assistants can be individual users or organizations that care about AI safety. Since LLMs learn and use the data they have been fed as well, issues in privacy and security arise as a valid ethical concern in the usage of various large language models. The purpose of such rules is to prevent any sensitive or unwanted data from reaching the LLM providers and to prevent any such data from reaching the end user as well.

\subsection{Types of rules}

There are three main types of rules that differ in their technical difficulty and in their strength. The user can use all or some of these to create fully customizable policies reflecting their ethical choices and privacy concerns.

The first one comprises \textit{static rules} that consist of predefined patterns that AI assistants use to identify and filter easily predictable sensitive information such as email addresses, Social Security numbers, phone numbers, and other \textit{personally identifiable information} (PII). For example, a regex or a similar natural language-processing (NLP) pattern recognition mechanism might detect or mask out PII to prevent unintentional exposure of private data to third-party LLM providers.

\textit{Natural-language rules} are expressed by the user in human-readable language and provide guidelines on what behavior and content to either encourage or avoid in conversations made with an LLM. For example, ``never mention any content inappropriate for children below the age of 12'' or ``avoid conversation about religion''. This type of rule is fully customizable and can cover a wide range of considerations, either from maintaining a polite discourse and avoiding offensive language to adhering to industry-specific regulations. These are different from system prompts\footnote{System prompts guide the way AI models interpret and respond to user queries. See e.g. \cite{prompts} for more details.} since they can be combined with them without altering the original system prompt. System prompts also refer to the AI output only, but natural-language rules can be used for user input as well.

In order not to fall into the same trap of using an LLM to mitigate responses from an LLM, two approaches can be taken here. First, a natural language rule can be enforced using natural language processing techniques without using LLMs. For example, various predefined lists of keywords and phrases can be used along with the user-defined description so that such lists can be avoided, along with various types of sentiment analyses and lexicon-based approaches. Another option is for an organization or a more technical user to host their own open-source versions of large language models such as Llama or Mixtral, leaving all of the user's data in their own hands.

\textit{Trained classifier rules} utilize machine-learning models to involve classifiers trained on labeled datasets. The user can start creating their own dataset by adding examples that are to be denied by the classifier and ones that are to be allowed. The user can upload their own datasets, use publicly available ones for fine-tuning, or use LLMs to generate synthetic few-shot examples that are similar to the ones chosen by the users. For instance, a classifier trained to detect urgency in medical assistance could prioritize responses to critical medical inquiries over general queries.

Non-technical users are likely to find natural-language rules particularly appealing due to their simplicity and accessibility. These rules allow users to express their ethical guidelines and preferences in plain, everyday language without needing to understand complex technical details or programming. For example, a business owner might easily set up rules like ``ensure all communications remain professional'' or ``avoid discussing sensitive topics like politics and religion'', without requiring any specialized knowledge of natural language processing or machine learning. However, specific trained classifiers, when provided with enough data, can offer a more nuanced and precise level of control over AI behavior. These classifiers can be tailored to recognize and respond to highly specialized or context-specific content, making them particularly valuable in industries where precision is critical.

\subsection{Policies}

Various rules can be combined into \textit{policies}. Rules within policies are evaluated sequentially by default to determine compliance. Static rules come first so that immediate matches based on easily predictable patterns can be detected upfront and filter out sensitive information. Next, natural-language rules are then enforced, influencing the conversation so that it aligns with the user-defined ethical norms and requirements. Lastly, trained classifier rules classify both inputs and outputs based on learned categories, which allows a more nuanced understanding and response generation.

This default sequence can be altered by the user to create a \textit{hierarchical chain of rules}, again emphasizing various ways a compliant system might be customized that depend on the end user's privacy and ethical preferences. All of the policies can be used for both the AI output and the user's input.

\subsection{Assistants}

A combination of input and output policies, along with preferred system prompts and action items, creates an \textit{AI assistant}. An \textit{input policy} governs how the user can behave and what questions they can ask, ensuring compliance with the organization's values and regulations. An \textit{output policy} follows the usual AI guardrail goal: govern how the AI assistant responds and interacts with its users, which ensures that responses align with user preferences or organizational standards.

When a rule within a policy is violated, this is labeled as inappropriate content so that the system can trigger a corresponding action, customized by the user. The user can choose what happens if a policy is violated, i.e., if a breach in one or more rules is detected. First, a redaction might happen, and the interaction between the AI system and the user may continue seamlessly, but the sensitive data is never sent to a third-party LLM provider. For example, a rule may detect the user providing their Social Security number or trade secrets, and such information may be redacted. Second, a blocking violation may occur, in which the interaction is completely stopped, and, in the case of organizations, there may be additional actions happening, such as sending a warning notification, logging the breach, or notifying a human in the loop. A warning message is sent to the user, informing them of the policy violation and providing context-specific feedback or instructions. 

Depending on the severity of the violation, the AI assistant may restrict further interaction until the issue is resolved or avoided. For instance, in cases of repeated or severe policy breaches, the assistant could be programmed to temporarily block the user's access, escalate the issue to a human moderator, or offer corrective guidance to ensure compliance moving forward. This dynamic and responsive approach to rule enforcement not only maintains the integrity of interactions but also fosters a safer and more controlled environment for all users involved.

Ultimately, the assistant's ability to dynamically enforce these policies ensures that AI-driven interactions remain aligned with the ethical standards and operational goals of the end user or an organization, while also providing the flexibility needed to adapt to evolving requirements and contexts.

\subsection{Framework summary}

\begin{figure}
    \centering
    \includegraphics[width=\columnwidth,height=0.8\textheight,keepaspectratio]{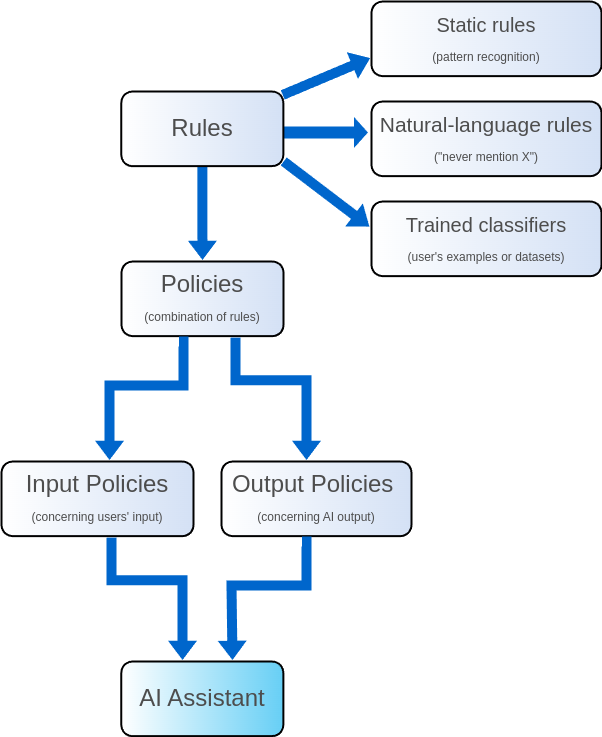}
    \caption{\small A hierarchical framework for managing and enforcing ethical standards in AI systems. This diagram illustrates the relationship between rules (static rules, trained classifiers, natural language rules), which combine to form policies. These policies are then integrated into AI assistants, ensuring that AI behavior aligns with specified ethical guidelines and operational goals.}
    \label{fig:hierarchy}
\end{figure}

This architecture, as illustrated in Figure \ref{fig:hierarchy}, provides a comprehensive framework for managing and enforcing rules in AI assistants, ensuring user trust and compliance across various domains. Deployment of AI guardrails (cf. \cite{nemo2023}, \cite{guardrailsai}, \cite{llamaguard}) represents a crucial step towards mitigating ethical concerns in AI systems.

However, these systems, despite the good intentions behind them, often operate as black boxes to end users, lacking transparency in how ethical decisions are made. In compliance with XAI, efforts have been made to enhance transparency in the guardrail portion of an AI system. The prototype in action is available publicly as an AI Trust Platform\footnote{Preamble AI Trust Platform (2024).} demonstrating these rules in practice.

\section{Ethical pluralism in AI design}

Ethical pluralism background acknowledges that diverse individuals and organizations may hold varying ethical perspectives and values, which underscores the need for customizable guardrails that can accommodate different ethical frameworks. By allowing users to easily define their values incorporated into rules and policies based on their specific ethical concerns, an AI system can be fully aligned not only to an abstract moral model but to diverse standards and norms.

Russell's \cite{russell2019} work argues that AI systems should be designed to align with human objectives (be ``human compatible''), rather than aimlessly pursuing some predefined goals that may conflict with human values. Russell's three principles include:
\begin{enumerate}
    \item \textit{The machine's only objective is to maximize the realization of human preferences.}
    \item \textit{The machine is initially uncertain about what those preferences are.}
    \item \textit{The ultimate source of information about human preferences is human behavior.}
\end{enumerate}

This approach is focused on the foundational aspects of AI ethics, while the guardrails approach in this paper emphasizes a \textit{practical} mechanism for implementing ethical standards in AI systems. The framework focuses on customizable rules and policies that allow users to define ethical guidelines specific to their values and contexts, which acknowledges the diverse ethical perspectives that are too vague in Russell's approach. The notion of ``human preference'' is already subject to questions of universality and relativism.

Both of these approaches emphasize \textit{transparency}, calling out for clear explanations of AI behavior and reasoning, and by allowing users to configure and monitor ethical rules and policies, they can easily make ethical decisions or readily see the consequences of such rules in practice (with belief-revision possibilities). Practical ethical AI agents should then follow a similar set of principles:
\begin{enumerate}
    \item \textit{The AI system's objective is to maximize the realization of user-specified ethical guidelines and values.}
    \item \textit{The machine and user are initially uncertain about what those preferences are, but the user has to have a transparent yet fully configurable way to communicate such preferences.}
    \item \textit{The ultimate source of information about the user's preferences is a set of rules and policies curated by the user, that can be analyzed and revised at will, according to the user's needs.}
\end{enumerate}

The user-centric ethical design of AI guardrails is pivotal in navigating the complexities of addressing value pluralism and the overall mistrust in AI systems that motivated the desire for explainable AI.

\section{Possible conflicts}

One possible issue is the question of conflicts between various guardrails inside a described framework. In addressing the complexities of guardrail conflicts within AI ethics systems, it is crucial to recognize the diverse scenarios in which such conflicts may arise and the mechanisms by which they can be managed. In this section, we will explore various cases of guardrail opposition, categorize them based on the nature of their conflict, and propose various strategies for their resolution.

\subsection{Case 1: complete and permanent opposition}
In this scenario, Guardrail A and Guardrail B are always in complete opposition, meaning their ethical vectors as policy combination representations are exact opposites, with a cosine similarity of $-1$. This kind of conflict represents a situation where two ethical directives are fundamentally irreconcilable. For instance, one guardrail might prioritize absolute privacy, while another emphasizes full transparency. Such a scenario could be identified through static analysis, flagging the inherent opposition before deployment, by analyzing the mathematical relationships or logical conditions set by these policies.

\textit{Variant I}: If Guardrails A and B are the only active guardrails, their mutual negation leaves the system without any ethical direction, resulting in an ``ethically blind'' state. This condition is particularly problematic as it disables the AI's ethical guidance entirely.

\textit{Variant II}: When other guardrails are active alongside A and B, the system may still function ethically as intended, relying on these additional guardrails. However, the persistent opposition between A and B may cause inconsistencies in ethical reasoning.

\textit{Variant III}: If all active guardrails are engaged in mutual negation, the system is left without any moral guidance, despite having multiple guardrails. This total negation creates a scenario where the AI operates without ethical constraints, which is highly undesirable.

\subsection{Case 2: permanent but limited disagreement}
Here, Guardrails A and B are generally opposed but not completely, with a cosine similarity close to $-1$, such as $-0.9$. This situation mirrors the persistent yet manageable disagreements seen in political discourse, such as those between major political parties. Although there is opposition, it allows for some degree of consensus through weighted averaging. Static analysis could identify these cases, enabling adjustments to achieve a balanced ethical stance.

\subsection{Case 3: Conditional Opposition}
In this case, Guardrails A and B are only sometimes in complete opposition, i.e., their cosine similarity is sometimes $-1$, depending on the specific context or input. The ethical vectors might align in some situations but conflict in others.

Variants I, II, and III from Case 1 apply here as well.

\subsection{Case 4: conditional but limited disagreement}
In this case, Guardrails A and B sometimes have opposing values but to a lesser degree, with a cosine similarity around $-0.9$. This situation is akin to a temporary political disagreement, where opposition exists but does not preclude finding common ground. Such conflicts are usually manageable within the system's existing framework, using weighted averaging to navigate the disagreement.

\subsection{Conflict Resolution Strategies}
To manage these conflicts, we consider the following strategies:

\begin{enumerate}
\item \textit{Weighted averaging system}. In most scenarios, especially in Cases 2 and 4, a weighted averaging system allows for nuanced ethical reasoning by considering the strengths of various guardrails. However, this approach struggles with Cases 1 and 3, where complete opposition can lead to ethical paralysis.

\item \textit{Strict order or hierarchy of precedence}. To address the limitations of weighted averaging, a strict order of precedence can be established, where higher-priority guardrails override others in cases of conflict. While this prevents total ethical blindness, it risks allowing weakly held opinions from high-precedence guardrails to dominate more strongly held positions from others, leading to potential ethical imbalances.

\item \textit{Hybrid approach: conditional precedence}. A hybrid approach could involve using weighted averaging as the default method but reverting to a strict order of precedence when mutual negation is detected (e.g., in Case 1/I, 1/III, 3/I, or 3/III). This system would handle temporary conflicts effectively while ensuring that permanent oppositions trigger alerts, warning users that the system is operating under constrained ethical guidance.

\item \textit{Contextual triggering}. The AI system can be designed to apply different guardrails based on the specific context or scenario. For example, in situations involving sensitive personal information, the privacy guardrail might be activated, overriding the transparency directive. Conversely, in situations requiring public accountability, the transparency guardrail could take precedence. This approach allows for a dynamic resolution of conflicts based on real-time analysis of the situation.

\item \textit{User resolution.} In cases where automatic resolution is challenging or where both guardrails are of equal importance, the system could flag the conflict for human intervention. Users or administrators can then manually decide which guardrail should take precedence in the given situation. This method is particularly useful in high-stakes environments where nuanced human judgment is necessary, invoking the \textit{human-in-the-loop} strategy often present in AI systems.
\end{enumerate}

Understanding and addressing guardrail conflicts is essential for the development of robust AI systems. While Cases 2 and 4 represent manageable, everyday ethical disagreements, Cases 1 and 3 pose significant challenges that require thoughtful design and conflict resolution strategies. By implementing a combination of weighted averaging and conditional precedence, we can create systems that maintain ethical integrity even in the face of complex, conflicting directives. This approach ensures that AI systems remain aligned with diverse human values, enhancing trust and reliability in their ethical behavior.

\section{Final remarks}

Our goal was to show that this approach of designing systems that prioritize user autonomy, transparency, and continuous improvement is a nudge in the right direction that promotes a collaborative approach where ethical standards evolve in response to feedback. Moving forward, continued research and innovation in AI ethics should aim to enhance transparency, accommodate diverse ethical perspectives, and allow the end users to navigate the ethical aspects of AI technologies effectively.

The introduction of customizable guardrails not only enhances the ethical robustness of AI systems but also fosters greater transparency and trust. As AI systems become increasingly integrated into decision-making processes, the ability to understand and influence the ethical reasoning of these systems becomes paramount. The framework proposed in this paper aims to bridge the gap between abstract ethical principles and practical implementation, offering a structured yet adaptable approach that can evolve alongside technological advancements and societal changes.

Moreover, the strategies for resolving conflicts between guardrails shed more light on the complexity of ethical decision-making in AI systems. By providing a range of resolution mechanisms, we ensure that AI systems can navigate ethical dilemmas in a manner that is both contextually appropriate and aligned with user expectations. This adaptability is particularly important in high-stakes environments where the consequences of AI decisions can have far-reaching impacts.

Looking forward, the continued refinement of these frameworks will be essential as AI systems are deployed in increasingly diverse and sensitive areas. Future research should focus on expanding the capabilities of guardrail systems, integrating more sophisticated context-awareness, and exploring new ways to involve users in the ethical governance of AI. Additionally, ongoing collaboration between AI developers, ethicists, and end users will be vital in ensuring that AI systems remain not only technically advanced but also ethically sound.

\section*{Declarations}




\subsection*{Data availability}
No datasets were generated or analyzed during the current study.

\subsection*{Competing interests}
The authors are affiliated with Preamble, whose AI Trust Platform is described in this paper.


\subsection*{Authors' contributions}
All authors contributed to the conception and design of the work. K\v{S} drafted the manuscript and led the engineering part. JM and JRC contributed to the development of the framework and the AI Trust Platform prototype. All authors read and approved the final manuscript.


\bibliography{bibliography}
\bibliographystyle{apalike}

\end{document}